\documentclass[a4paper,twoside,reqno]{bjp}
\usepackage{graphicx}
\usepackage{cite}
\usepackage{amssymb,amsmath,amscd,amsthm}
\usepackage{times}

\usepackage[bookmarks=false]{hyperref}
\hypersetup{%
    colorlinks=true,        
    linkcolor=blue,          
    citecolor=blue,         
    urlcolor=blue           
    }

\usepackage{geometry}
\allowdisplaybreaks

\begin{document}

\title{Hyperspherical few-body  model calculation for cascade $\Xi$ exotic hypernuclei with reference to nuclear drip lines.}

\runningheads{Hyperspherical few-body  model calculation for cascade $\Xi$ exotic hypernuclei....}{Md. A. Khan et al.}

\begin{start}{%
\author{Md. A. Khan}{1},
\author{M. Alam}{1},
\author{M. Hasan}{1},
\author{S. H. Mondal}{1}

\address{Department of Physics, Aliah University, \\ IIA/27, Newtown, Kolkata-700160, India}{1}

\received{Day Month Year (Insert date of submission)}
}

\begin{Abstract}
In this paper, we presented the ground state energies calculated for some single and double cascade hyperon(${\Xi}^{-}$) hypernuclei. For hypernuclei of the type $_{\Xi}^{A_c+1}$X we applied $\Xi+^A_c$X two-body model and for hypernuclei of the type $_{\Xi\Xi}^{A_c+2}$X we adopted $\Xi+\Xi+^A_c$X three-body model. A two-term Gaussian (Isle type) potential with adjustable parameters is chosen as the $\Xi$-nucleus interaction, while a two-term Yukawa-type potential has been chosen for the $\Xi\Xi$ pair. In the three-body model calculation, we adopted the hyperspherical harmonics expansion (HHE) formalism. The resulting two- and three-body Schr{\"o}dinger equations subject to appropriate boundary condition's have been solved numerically to get the ground-state energies and wavefunctions. Computed values are in excellent agreement with the observed ones as found in the literature. Analysis of the calculated observables indicates that the hyperons stay close to the core-nucleus. Hypernuclei formed as a result of absorption of one or more $\Xi$-hyperon(s) by a normal nucleus (bound or unbound) are found to have more stability than those formed as a result of absorption of one or more $\Lambda$ hyperon(s) by the same nucleus. This fact flashes towards the possibility of the existence of a valley of strange hypernuclei near the drip lines.
\end{Abstract}

\begin{KEY}
 Hyperon, Hypernuclei, exotic-nuclei, cascade hyper-nuclei, hyperon-nucleon potential, hyperspherical harmonics.
\end{KEY}
\end{start}

\section{Introduction}
Atomic nuclei containing at least one strange hyperon (viz. $\Lambda, \Sigma, \Xi$, etc) in their constituents are called hypernuclei. Nuclear physics of strange particle is an exciting and vibrant field of research, dealing with the production of multi-strange hypernuclei (sometimes called strangelets) in the laboratory through relativistic heavy-ion collisions. When two heavy ions collide at very high energy, constituent nucleons of the colliding ions undergo multiple collisions. Each independent nucleon-nucleon collision is capable of producing a strange quark-antiquark pair (i.e., $s \bar{s}$ pair). As the quark structure of proton (uud), neutron (udd), and $\Xi^{-}$ (dss) are well-established facts, interaction among hyperons and nucleons may give a very good starting point to extract information on strong interactions through investigation of hypernuclei. An interesting type of them is the cascade hypernuclei, which are formed when one or more nucleon(s) of ordinary nuclei are replaced by one or more cascade (eg., $\Xi$) hyperon(s). Some of the experimentally observed cascade hypernuclei are $_{\Xi^-}^{8}$He, $_{\Xi^-}^{15}$C, $_{\Xi^-}^{17}$O etc. The existence of cascade hyperon $\Xi^-$ was first inferred in the works of the Manchester group of scientists- Armenmteros R., Barker K. H., Butler C. C., Cachon A and York C. M. in the early fifties \cite{armenteros}. Existence of a neutral component of which was predicted by Gell-Mann M. 1956 \cite{gell} based on strangeness theory. The prediction was later confirmed in experiments of the Lawrence Radiation Laboratory group led by Alvarez L W 1959 \cite{alvarez}. This particle was also observed at BNL, as an intermediate product of $\Omega^-$ decay, during the mid-sixties, \cite{barnes}. The possibility of the formation of cascade hypernuclei was first reported by Barkas et al 1959 \cite{barkas}. Reports on the possible bound states of $\Xi-$hypernuclei have been published by several authors including Wilkinson et al. 1959 \cite{wilkinson}, Bechdolff et al. 1968 \cite{bechdolff}, Catala et al. 1969 \cite{catala}, and Mondal et al. 1979 \cite{mondal}. In their 1983 works on the formation of $\Xi$-hypernuclei, Dover and Gal reported some important pieces of information which can be seen in the literature \cite{dover1}. Experimental data on some $\Xi$-nucleus interaction and bound state properties of $_{\Xi}^{13}$C have been reported by Aoki et al. 1995\cite{aoki}. Fukuda et al. 1998 \cite{fukuda} published data on $\Xi$-nucleus potential based on their scattering and structure investigations. Khaustov et al. 2000 \cite{khaustov} reported the event of production of $_{\Xi}^{12}$Be hypernuclei and useful data on $\Xi$-nucleus potential strength. Nakazawa et al. 2015 \cite{nakazawa} reported a deeply bound state of the ${\Xi^-}^{14}$N hypernuclei having a $\Xi-$ separation energy of $4.38 \pm 0.25$ MeV. 
In addition to the above experimental activities, theoretical data on $\Xi$-hypernuclei obtained from momentum transfer reactions are also reported by Dover 1980 \cite{dover}, Dover, and Gal 1983 \cite{dover1}. Shoeb and Khan 1984 \cite{shoeb} published theoretical {\it ab-initio} data on binding energies of some light $\Xi$-hypernuclei. Similar theoretical results have also been reported by Lalazissis et al. 1989 \cite{lalazissis}. Yamaguchi et al. 2001 \cite{yamaguchi} used $\Xi N$ one boson exchange potential (OBEP) in the G matrix theory approach for studying bound states of $\Xi-^{11}$B hypernucleus. Hiyama et al. 2008 \cite{hiyama} performed structure calculation adopting a microscopic few-body cluster model using the Gaussian expansion method. Yamamoto et al. 2010 \cite{yamamoto} investigated bound states in some light and heavy $\Xi$-hypernuclei adopting G-matrix theory. Shyam et al. 2012 \cite{shyam} used an effective Lagrangian model approach to study cross-sections for the production of $\Xi^-$particle hole states and corresponding binding energies of $\Xi^-$ states in $_{\Xi^-}^{12}$Be, $_{\Xi^-}^{28}$Mg, etc. Sun et al. 2016 \cite{sun} theoretically identified the formation of $_{\Xi}^{15}$C state during capture of $\Xi^-$ by $^{14}$N target, using relativistic mean-field (RMF) and Skyrme-Hartree-Fock (SHF) models. Filikhin et al. 2017 \cite{filikhin1} have implemented configuration-space Faddeev equation approach to study binding mechanism of three-body systems containing hyperon(s), nucleon(s) and nuclide(s). Very recently Liu et al. 2018 \cite{liu} studied the separation energies of hyperon(s) in hypernuclei $_{\Xi^-}^{41}$K, $_{\Xi^0}^{41}$Ca etc in the relativistic mean-field approach.

Also, hypernuclei exhibit some interesting features-they are self-bound quantum many-body systems embodied of three different types of fermions-proton, neutron, and strange hyperon. Being non-identical fermion, an unpaired hyperon does not suffer from Pauli blocking by the nucleons. As a result, a hyperon can penetrate deep into the nuclear interior and occupy any of the allowed energy orbits, giving rise to a phenomenon known as {\bf glue-like} role of hyperon. Thus hyperons can generate deeply bound hypernuclear states as reported by Hiyama et al. 2015 \cite{hiyama1}, Liu et al. 2018 \cite{liu}, and Xia et al. 2019 \cite{xia}. Buyukcizmeci et al. 2013 \cite{buyukcizmeci} noted that nucleon separation energies in hypernuclei are relatively higher than those in normal nuclei due to coupling between hyperons and nucleons inside the hypernuclei. This interesting phenomenon may provide the scientific community a noble opportunity to investigate the formation and structure of normal exotic nuclei beyond the drip lines. Hyperons may produce some other interesting effects in nuclear systems as an impurity such as shrinkage of nuclear size  Hiyama et al. 2010 \cite{hiyama2}; hyperon resonance states Ren et al. 2017 \cite{ren}; nuclear shape deformation Isaka et al. 2013 \cite{isaka}, Lu et al. 2014 \cite{lu}; modification of cluster structure by penetrating deep inside the target nucleus Hagino et al 2013 \cite{hagino}, a shift of neutron drip line towards the neutron-rich side Zhou et al. 2008 \cite{zhou} and occurrence of the skin or halo structure of the nucleon and hyperon Lu et al. 2003 \cite{lu1}. These strange nuclei have a lifetime significantly longer than the typical reaction times Danysz and Pniewski 1953 \cite{danysz}, and they can provide an extreme test for the models of nuclear structure, to study the hyperon-nucleon,  hyperon-hyperon interactions, and the strange matter properties. Also, the discovery of such strange nuclei has opened an avenue for the study of significant events not only in particle physics but in nuclear-astrophysics as well. Hyperons form the main constituent in the core of many stars and appear in the neutron stars when their matter density becomes approximately two times the density of matter in ordinary nuclei. The hyperon-nucleon and hyperon-hyperon interactions have adequate demands to the solution of the equation of state (EoS) describing a neutron star and important to calculate the maximum mass and size of the neutron stars. Also, hyperons provide additional stability and their interactions regulate the cooling behavior of the massive neutron stars Schaffner 2008 \cite{schaffner} and Belyaev et al. 2008 \cite{belyaev}.

In the literature we found some scattered theoretical and experimental data on the single $\Xi$-hypernuclei, such as those by Bando et al. 1990 \cite{bando}, Feliciello et al. 2015 \cite{feliciello}. However, data on the corresponding double-$\Xi$ hypernuclei are very scarce. This fact insisted us to investigate the possible bound states in the single- and the corresponding double-$\Xi$ hypernuclei using the same set of hyperon-nucleus interaction. The $\Xi\Xi$ interaction is chosen to keep in mind the prediction of the extended softcore Nijmegen potential model (ESC08c), that no $\Xi\Xi$ bound state to exist, reported by Garcilazo and Valcarce 2016 \cite{garcilazo}.

This work is motivated by the recent experimental activities worldwide to unveil many interesting properties of this exotic kind of hyperon playing a very crucial role near the nuclear drip line. Here we carry out an ab-initio solution of Schr\"{o}dinger equation to calculate energy and some relevant parameters of single and double $\Xi$ hypernuclei in the framework of few-body (two and three-body) model adopting hyperspherical harmonics expansion method (HHEM). As already discussed, many exotic phenomena occur near the drip line and {\bf formation of bound exotic hyper-nuclei via capture of strange hyperons by unbound or very weakly bound normal nuclei near the drip line} is one of the most interesting events among them. Here we considered hypernuclei of mass number A = 6 \& 16 assumed to be formed via absorption of two $\Xi^-$ hyperons by $^4$He and $^{14}$C respectively. It is known that $^5$He is unbound and $^6$He has small two-neutron separation energy (less than 1 MeV). One of the interesting features of $^6$He is that it is a Borromean halo nucleus because none of the binary subsystems of $^6$He three-body ($^4$He+n+n) system are bound which resembles the Borromean structure of three interlocked rings which separate automatically if any one of the rings is removed. Thus, the existence of the bound state of $_{\Xi^-}^5$He in comparison to the non-existence of $^5$He bound state or the enhancement in the binding energy of $_{\Xi\Xi}^6$He over the binding energy of $^6$He could validate our forgoing remarks. For this purpose, we adopt a few-body (two- and three-body) model for the single and double- $\Xi$ hypernuclei respectively. In our model, we treat the relatively heavier nucleus as the core and the remaining one or two $\Xi$ hyperon(s) as the valence particle(s). For example, in $_{\Xi\Xi}^{6}$He, we treat $^4$He as the core nucleus and two $\Xi$ hyperons as valence particles. In our scheme, we will first try to reproduce the observed data for single -$\Xi$ hyperon hypernuclei by adjusting the parameter of the effective core-hyperon potential. In the next step, we use the same $\Xi$-nucleus potential parameters together with chosen $\Xi\Xi$ potential to solve three-body Schr\"{o}dinger equation to predict the ground state energy and wavefunction of the corresponding double $\Xi$-hypernuclei. Computed wave-functions will be used to compute 3body matter radii to see whether there is any shrinkage effect in the size of target nuclei due to the absorption of $\Xi$ hyperon(s). The paper is organized as follows. In section 2, we describe the HHE method for a three-body system consisting of two identical particles, section 3 will be devoted to results and discussion. Finally, in section 4, we shall draw our conclusions.

\section{Hyperspherical Harmonics Expansion (HHE) Method}
\begin{figure}[htb]
\centerline{\includegraphics[width=0.65\textwidth]{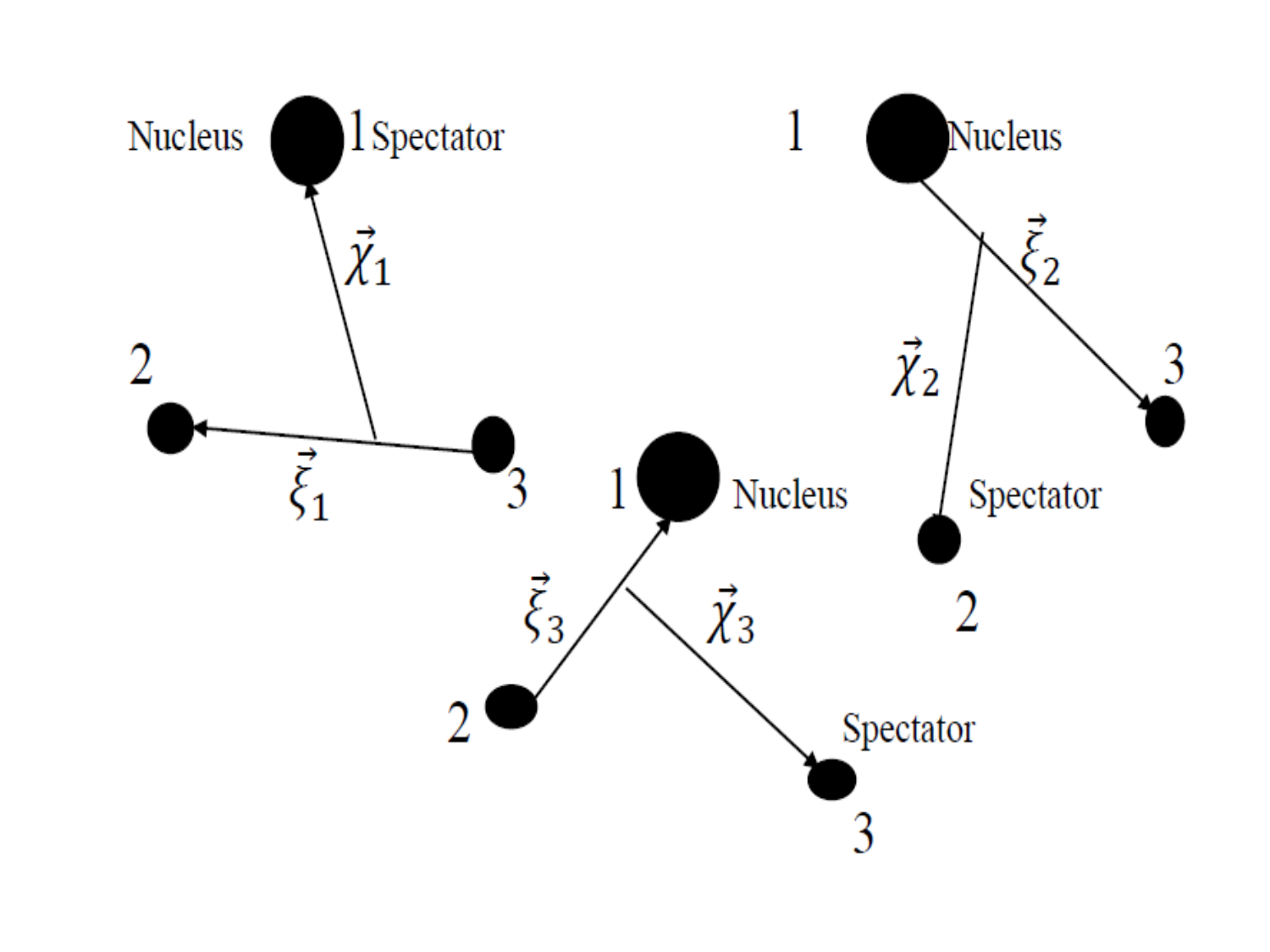}}
\caption[]{Label scheme and choice of Jacobi coordinates for a general three-body system in the partition $\lq\lq i$", [i=1,2,3 - cyclic].}
\label{f01}
\end{figure}

The HHE method adopted here is essentially an exact method except for an eventual truncation of the expansion basis due to computer memory limitations, in which calculation can be performed for any desired precision by gradually increasing the size expansion basis \cite{khan1} and checking the corresponding convergence trend. The method has been used successfully in our previous work on atomic and nuclear systems \cite{khan1, khan2, khan3, khan4}. In this approach, the labeling scheme and choice of Jacobi coordinate for a general three-body system consisting of a relatively heavier core nucleus and two valence $\Xi$-hyperons are shown in Figure~\ref{f01}.
The Jacobi coordinates corresponding to the partition $\lq\lq i$" in which particle labelled $\lq\lq i$" is the spectator and those labelled as $\lq\lq j$" and $\lq\lq k$" form the interacting pair are defined as:
\begin{gather}
\left.  \begin{array}{ccl}
 \vec{\xi_{i}}& = &\left[ \frac{m_{j} m_{k}M}{m_{i}(m_{j}+m_{k})^{2}}
\right]^{\frac{1}{4}} (\vec{r_{j}} - \vec{r_{k}})\\
   \vec{\chi_{i}} & = & \left[ \frac{m_{i} (m_{j}+m_{k})^{2}}
{m_{j} m_{k} M}\right]^{\frac{1}{4}} \left( \vec{r_{i}} - \frac{m_{j}
\vec{r_{j}} + m_{k} \vec{r_{k}}}{ m_{j} + m_{k}} 
 \right)
  \end{array}  \right\}\label{eq01}
\end{gather}
where $\vec{R} = \sum_{i=1}^3m_{i}\vec{r_{i}}/M$ is the centre of mass coordinate, $M = (m_i+m_j+m_k)$  is the total mass of the system, and sign of $\vec{\xi_{i}}$ is determined by the condition that ($ i, j, k $) form a cyclic permutation of (1, 2, 3). Hyperspherical variables associated with the Jacobi coordinates having a more elaborate description in Khan, 2012 \cite{khan1} are:
\begin{gather}
\xi_{i}  =  \rho \cos \phi_{i};\: \chi_{i}  =  \rho \sin \phi_{i}\label{eq02}
\end{gather}
where $\rho =\sqrt{ \xi_{i}^{2} + \chi_{i}^{2}}; \phi_{i} = \tan^{-1}(\chi_i/\xi_i)$. The motion of the three-body system in relative coordinates are then described by the Schr\"{o}dinger's equation
\begin{gather}
\left[ - \frac{\hbar^{2}}{2\mu}\left\{ \frac{\partial^2}{\partial\rho^2}+ \frac{5}{\rho} \frac{\partial}{\partial\rho}+
\frac{\hat K^{2}(\Omega_{i})}{\rho^{2}} \right\}+ V (\rho, \Omega_{i}
) - E \right] 
\psi (\rho, \Omega_{i} ) \:=\: 0 \label{eq03}
\end{gather}
where  ${\mu\: =\: \left[ \frac{m_{i} m_{j} m_{k}}{M} \right]^{\frac{1}{2}}}$ is an effective mass parameter, $V (\rho, \Omega_{i})$ = $ V_{jk} + V_{ki} + V_{ij} $ is the total interaction potential, and $\Omega_{i} \rightarrow \{ \phi_{i}, \theta_{\xi_{i}}, \phi_{\xi_{i}}, \theta_{\chi_{i}}, \phi_{\chi_{i}} \}$ represents hyper-angles. The wave function $\psi(\rho, \Omega_{i})$, in any chosen partition (say in the partition $\lq\lq i$") is expanded as
\begin{gather}
\psi(\rho, \Omega_{i}) = \sum_{K \sigma_{i}}\rho^{-5/2}W_{K \sigma_{i}}   
  (\rho) {\Theta}_{K \sigma_{i}}(\Omega_{i})\label{eq04}
\end{gather}
In Eq. (\ref{eq03}), square of hyper angular momentum operator $\hat K^{2}(\Omega_{i})$ satisfies an eigenvalue equation \cite{khan1}
\begin{gather}
\hat K^{2}(\Omega_{i}) {\Theta}_{K \sigma_{i}}(\Omega_{i}) = K(K + 4)  {\Theta}_{K \sigma_{i}}(\Omega_{i}) \label{eq05}
\end{gather}
corresponding to the hyperspherical harmonics (HH) $\Theta_{K \sigma_{i}}(\Omega_{i})$, $K=2n_i+ l_{\xi_{i}} + l_{\chi_{i}}, \:n_i$ being non-negative integer; $\sigma_{i}=\{ l_{\xi_{i}}, l_{\chi_{i}}, L, M \}$, angular momenta $l_{\xi_{i}}, l_{\chi_{i}}$ for $\vec{\Xi_i}, \vec{\chi_i}$ motions are coupled to define the total orbital angular momentum $\vec{L}=\vec{l_{\xi_{i}}} + \vec{l_{\chi_{i}}}$, M is the projection of $\vec{L}$. Substitution of expansion Eq.~(\ref{eq04}) in Eq.~(\ref{eq03}) and use of the orthonormality of HH lead to an infinite set of coupled differential equations (CDE) in $\rho$
\begin{gather}
\left[ -\frac{\hbar^{2}}{2\mu} \frac{d^{2}}{d\rho^{2}}
+\frac{\hbar^2}{2\mu}\frac{(K+\frac{3}{2})(K+\frac{5}{2})}{\rho^{2}} - E \right] 
W_{K \sigma_{i}}(\rho)+ \nonumber \\ \sum_{K^{\prime} \sigma_{i}^{\prime}} M_{KK^{\prime}}^{\sigma_i\sigma_{i}^{\prime}}
W_{K^{\prime} \sigma_{i}^{~\prime}}(\rho) \:= \: 0.\label{eq06}
\end{gather}
where
\begin{gather}
M_{KK^{\prime}}^{\sigma_i\sigma_{i}^{\prime}} = <K \sigma_{i}
\mid V(\rho, \Omega_{i}) \mid K^{\prime} \sigma_{i}^{\prime}
> =  \int
{\Theta}_{K \sigma_{i}}^{*}(\Omega_{i}) V(\rho, \Omega_{i}) {\Theta}_{K^{\prime} 
 \sigma_{i}^{~\prime}}(\Omega_{i}) d\Omega_{i}\label{eq07}
\end{gather}
represents the coupling potential matrix elements.
Expansion in Eq.~(\ref{eq04}) is truncated to a finite set due to practical limitations, resulting in a finite set of CDE. The imposition of symmetry and conservation conditions further reduces the expansion basis to a solvable set of CDE. After the evaluation of coupling potential matrix elements following the prescription of Khan, 2012 \cite{khan1}, the set of CDE Eq.~(\ref{eq07}) is solved numerically subject to proper boundary conditions to get the energy $E$ and the partial waves $W_{K\sigma_{i}}(\rho)$.

\section{Results and discussions}
For the core-$\Xi$ subsystem we used a two-range Gaussian Isle-type potential of Filikhin et al. 2008 \cite{filikhin}, 2017 \cite{filikhin1} of the form
\begin{gather}
V_{core-\Xi}(r)=V_{rep}\exp(-(r/\beta_{rep})^{2})-V_{att}\exp(-(r/\beta_{att})^{2}) \label{eq08}
\end{gather}
orginally derived for $\alpha\Xi$ system having parameters $V_{rep}$ = 450.40 MeV, $\beta_{rep}$ = 1.269 fm, $V_{att}$ = 404.90 MeV and $\beta_{att}$ = 1.41 fm. For other $\Xi$-hypernuclei an effective hyperon-nucleous potential is obtained by adjusting the depth parameter of the attractive component $V_{att}$ of the potential to reproduce the observed single-$\Xi$ separtion energy $(B_{\Xi})$ of the core-$\Xi$ subsystems leaving other parameters unaltered. For double $\Xi$-hypernuclei, we used Nijmegen extended softcore (ESC08c) Yukawa-type two-term potential for the $\Xi\Xi$ pair found in the works of Garcilazo et al. 2016 \cite{garcilazo}, Filikhin et al. 2017 \cite{filikhin1} and references therein:
\begin{gather}
V_{\Xi \Xi}(r)=(-155.0\exp(-1.75r)+490.0\exp(-5.6r))/r \label{eq09}
\end{gather}
A Coulomb interaction is added for the charged pair of interacting particles
\begin{gather}
V_{ij}^{(C)}(r_{ij}) = \left\{ \begin{array}{l}
1.44 \left( \frac{Z_iZ_j}{2R_c}\right) \left( 3 - \frac{r_{ij}^2}{R_c^2} \right) \:\: when \:\: r_{ij}\leq R_c\\
1.44 \left(\frac{Z_iZ_j}{r_{ij}}\right) \:\: when \:\: r_{ij}> R_c\\
\end{array} \right.\label{eq10}
\end{gather}
where $R_c = R_0 A^{1/3}$ is the radius of the core nucleus. In a hypernucleus since the core nucleus contains only nucleons and no $\Xi$ hyperon, no Pauli symmetry requirements under exchange of valance hyperon(s) and nucleons occur. However, symmetry requirement arises due to (a) antisymmetrization of the nuclear wave function under the exchange nucleons, which is suppressed in the choice of the core as a building block and (b) antisymmetrization of the three-body wave function under exchange of the valence hyperons have been correctly incorporated by restricting $l_{\xi_i}$ values. Most of the observed double-hyperon hypernuclei have total angular momentum J = 0, and even parity in their ground state. For spin-less nuclear core (i.e., $s_1 = 0$), total spin ($S = s_1 + s_2 + s_3$) of the three-body system has two allowed values either 0 or 1, since hyperons are fermions having odd-half spins. Since, J = 0, then for S = 0, L = 0; and for S = 1, L = 1. Hence, the spectroscopic notation for the ground state of double-hyperon hypernuclei are $^1S_0$ and $^3P_0$ corresponding to J = 0, S = 0, L = 0 and J = 0, S = 1, L = 1 respectively.

The core nucleus being spinless, $S = 0$ corresponds to zero total spin of the valence hyperons (i.e. $S_{23} = 0$), hence the spin part of the three-body wave function is antisymmetric under exchange of spins of two $\Xi$ hyperons, leaving the spatial
part of the wavefunction symmetric under exchange of their spatial coordinates. Under pair exchange operator $P_{23}$, which interchanges particles 2 and 3, $\xi_i\rightarrow -\xi_i$ and $\chi_i$ remains unchanged. Thus $P_{23}$ acts like the parity operator for the pair (23) only. If we consider two valance $ \Xi $-hyperons in spin singlet state (which is spin anti-symmetric), the space part of their wave function must be symmetric under $P_{23}$. This restricts $l_{\xi_i}$ to take even values only. Again, for singlet spin state, total orbital angular momentum $L=0$, so  $ l_{\xi_{i}}=l_{\chi_{i}}=$ even integer. Finally, since $K = 2n_{i}+l_{\xi_{i}}+l_{\chi_{i}}$, $n_i$ being a non-negative integer, K must be even and
\begin{gather}
l_{\xi_{i}} = l_{\chi_{i}} = 0, 2, 4, .....,\frac{K}{2} [or \: (\frac{K}{2} -1)] \: \\for \: \frac{K}{2}\: being \:even \:or \:odd \: respectively.\nonumber\label{eq11}
\end{gather}
For $S = 1$, two $\Xi$-hyperons will be in the symmetric spin triplet state $S_{23} = 1$ and the corresponding space wavefunction must be antisymmetric under $P_{23}$, allowing only $l_{\xi_i}$ = odd  integral values. As, the total orbital angular momentum $L = 1, |(l_{\xi_i}-1)|\leq l_{\chi_i} \leq (l_{\xi_i}+1)$, but parity conservation allows $l_{\chi_i} = l_{\xi_i}$ only. Thus $K=2n_{i}+l_{\xi_{i}}+l_{\chi_{i}}$, with non-negative integer $n_i$, allows only even values of $K$, so
\begin{gather}
l_{\xi_{i}} = l_{\chi_{i}} = 1, 3, 5, ....., \frac{K}{2} [or \: (\frac{K}{2} -1)] \: \\for \: \frac{K}{2}\: being \:odd \:or \:even \: respectively. \nonumber\label{eq12}
\end{gather}
In the practical calculation, the HH expansion basis given by Eq.~(\ref{eq04}) is truncated to a maximum value $(K_{max})$ of K and $ 0\le (l_{\xi_{i}}+l_{\chi_{i}})\le K $. For each allowed $ K\le K_{max}$ with K = even integers, all allowed values of $l_{\xi_{i}}$ are included. The even values of  $l_{\xi_{i}}$ correspond to L = 0, S = 0 while the odd values of  $l_{\xi_{i}}$ correspond to L = 1, S = 1. This truncates Eq.~(\ref{eq06}) to a set of N coupled differential equations gven by
\begin{gather}
N = \frac{1}{8}(K_{max}^2+6K_{max}+8) [when\: \frac{K_{max}}{2} \: is \: even] \nonumber\\= 
\frac{1}{8}(K_{max}+2)^2 [when \:\frac{K_{max}}{2} \: is \: odd]  \label{eq13}
\end{gather}
It follows from Eq.~(\ref{eq13}), that the number of CDE’s increases in quadratic form with $K_{max}$, hence in practice Eq.~(\ref{eq06}) is truncated to a managable size depending on the available computer facility and the resulting set of CDE's are solved using hyperspherical adiabatic approximation (HAA) \cite{das3}. The calculated single-$\Xi$ separation energies for few single-hyperon hypernuclei together with the depth of the attracticve component of effective hyperon-nucleon (YN) potential are presented in Table~\ref{t01}. Experimental and theoretical data found in the literature for these systems are also presented in Table~\ref{t01} for comparision. \\

\begin{table}[htbp]
\caption[]{Parameters of the attractive component of the core-$\Xi$ potential and corresponding single-$\Xi$ separation energy ($B_{\Xi}$) for few different single- $\Xi$ hypernuclei.}\small\smallskip

\tabcolsep=3.6pt
\begin{tabular}{@{}l|cc|ccr@{}}
\hline
\hline
&&&&&\\[-8pt]
\textbf{Systems}&\multicolumn{2}{c}{\textbf{Potential parameters}}\vline&\multicolumn{3}{c}{\textbf{$B_{\Xi}$ (MeV)}}\\
\cline{2-6}
&$V_{att.}$ &\textbf{$ \beta_{att.} $ (fm)} & \textbf{Present work} & \textbf{Other's work}&\textbf{Expt}.\\
\hline
&&&&&\\[-8pt]
$_{\Xi^{-}}^{5}$He&404.900&1.41&3.3396 &3.61 \cite{filikhin}    &\\
      $ _{\Xi^{-}}^{13}$C&381.450&1.41&3.7030&0.940 \cite{yamaguchi} &3.70$^{+18}_{-19}$ \cite{aoki}\\
                                        &&&&14.4 \cite{filikhin} & 18.1$\pm$3.2 \cite{mondal}\\
                                        $ _{\Xi^{-}}^{15}$C&387.961&1.41&4.3894&4.4 \cite{sun}&4.38$\pm$0.25 \cite{nakazawa} \\
                   &    & &&14.4 \cite{lalazissis}& 16.0$\pm$4.7 \cite{bechdolff}\\
$ _{\Xi^{-}}^{17}$O&442.100&1.41&16.0320&16.5 \cite{lalazissis}& 16.0$\pm$5.5 \cite{bechdolff}\\
$ _{\Xi^{-}}^{89}$Sr&417.670&1.41&21.3030&21.2 \cite{yamamoto}&\\
\hline
\hline
\end{tabular}
\label{t01}
\end{table}
Calculated double-$\Xi$ separation energies for the corresponding double-hyperon hypernuclei forming three-body systems are also presented in Table~\ref{t02}. \\

\begin{figure}[htb]
\begin{minipage}{12.5pc}
\centerline{\includegraphics[width=12.5pc,  height=12.5pc]{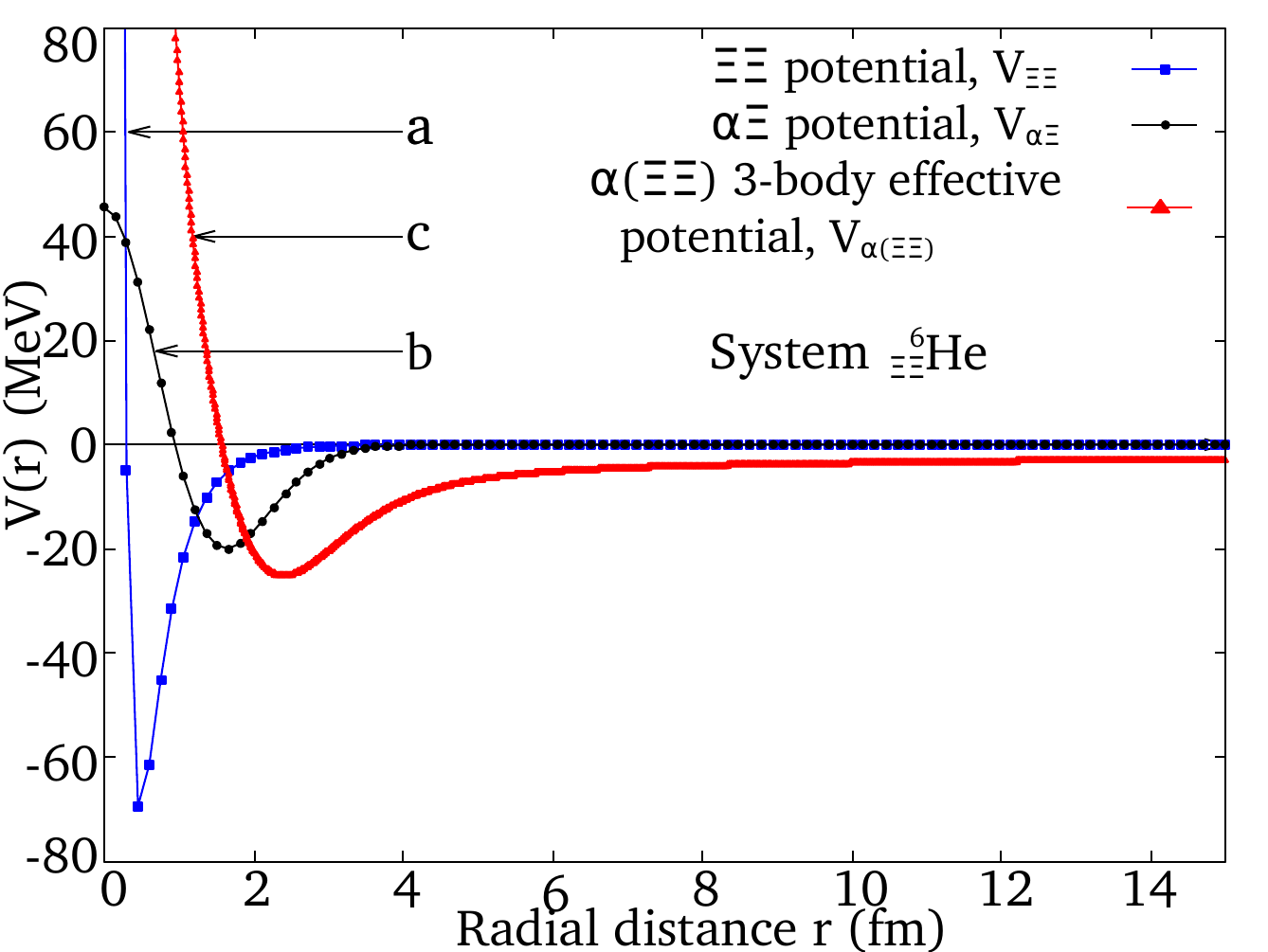}}
\caption{}{Representative plot of interaction potentials for $_{\Xi\Xi}^6$He: (a) $\Xi\Xi$ potential, $V_{\Xi\Xi}$; \cite{filikhin1} (b) $\alpha\Xi$ potential, $V_{\alpha\Xi}$;  and (c) $\alpha(\Xi\Xi)$ 3-body effective potential, $V_{\alpha(\Xi\Xi)}$.}
\label{f02}
\end{minipage}\hspace{0.5pc}
\begin{minipage}{12.5pc}
\includegraphics[width=12.5pc,  height=12.5pc]{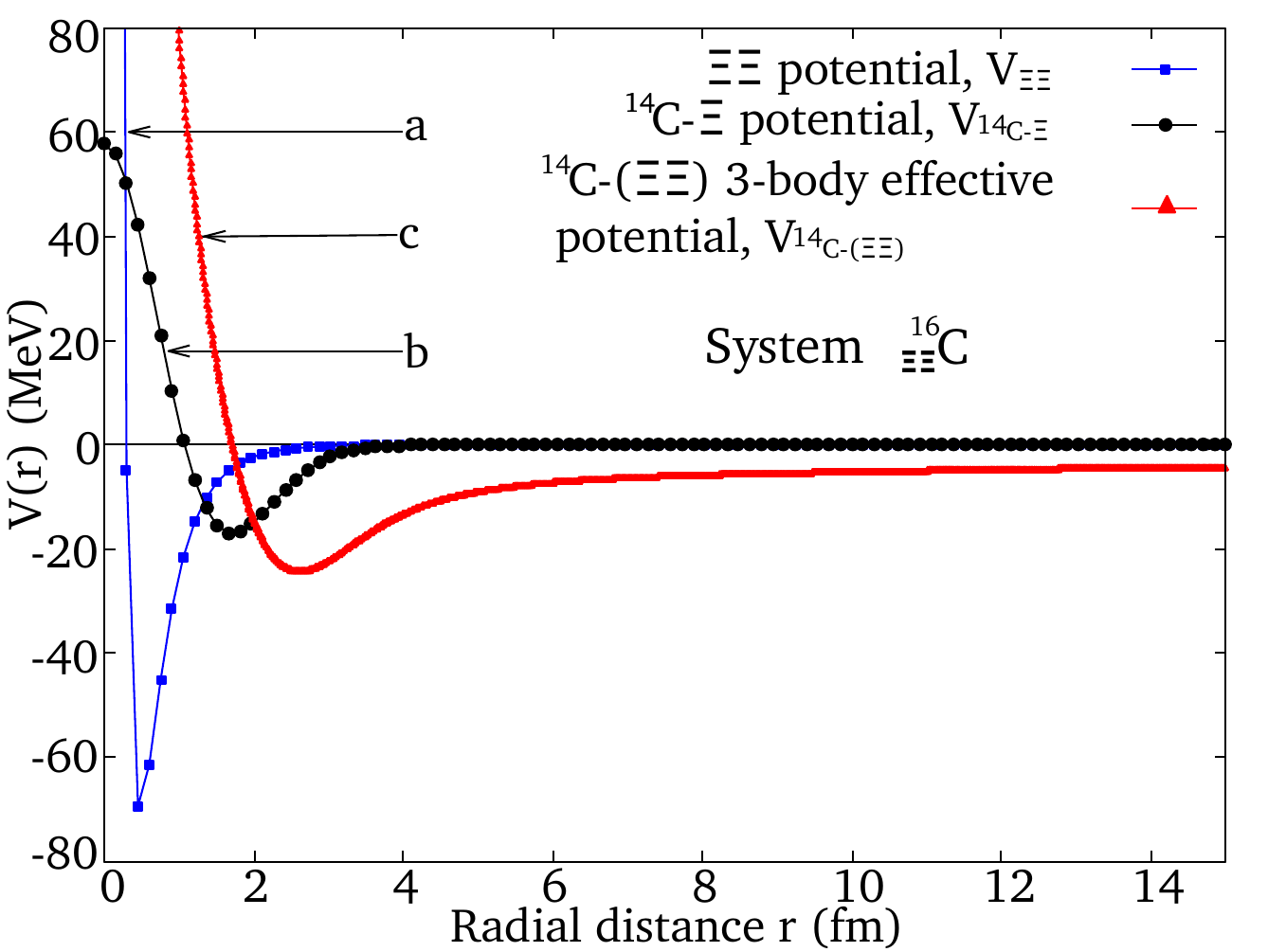}
\caption{}{Representative plot of interaction potentials for $_{\Xi\Xi}^{16}$C: (a) $\Xi\Xi$ potential, $V_{\Xi\Xi}$; \cite{filikhin1} (b) $^{14}$C$-\Xi$ potential, $V_{^{14}C-\Xi}$; (c) $^{14}$C-$(\Xi\Xi)$ 3-body effective potential, $V_{^{14}C-(\Xi\Xi)}$.}
\label{f03}
\end{minipage}\hspace{0.5pc}
\end{figure}

\begin{table}[htbp]
\caption[]{Calculated double-$\Xi$ separation energy $(B_{\Xi\Xi})$, $\Xi\Xi$ bond energy $(\Delta B_{\Xi\Xi})$ and geometrical observables for some double-$\Xi$ hypernuclei.}\small\smallskip

\tabcolsep=3.6pt
\begin{tabular}{@{}lccccc@{}}
\hline
\hline
&&&&&\\[-8pt]
\textbf{Systems}&\textbf{$_{\Xi\Xi}^{6}$He}&\textbf{$ _{\Xi\Xi}^{14}$C}&\textbf{$ _{\Xi\Xi}^{16}$C}&\textbf{$ _{\Xi\Xi}^{18}$O}&\textbf{$ _{\Xi\Xi}^{90}$Sr}\\
\hline
&&&&&\\[-8pt]
$B_{\Xi\Xi}$(MeV)&10.3149&9.9128&11.6269&38.5102&47.5565\\
$ \Delta B_{\Xi\Xi}$(MeV)&3.6357&2.5068&2.8479&6.4460&4.9505\\
$R_{3B}$&2.1341&2.5865&2.6589&2.6514&4.8301\\
$R_{c\Xi}$&2.5877&2.8587& 2.6970&1.8221&2.0752\\
$R_{\Xi\Xi}$&3.1432&3.5606&3.3506&2.2540&2.5436\\
$R_{(\Xi\Xi)c}$&2.0550&2.2367&2.1136&1.4318&1.6399\\
$R_{c}^{CM}$&0.8525&0.44253&0.3541&0.2144&0.0509\\
$R_{\Xi}^{CM}$&1.9796&2.5398&2.4295&1.6590&2.0353\\
$\eta$& 0.3564&0.3316&0.3299&0.3198&0.3167\\
\hline
\hline
\end{tabular}
\label{t02}
\end{table}
\noindent{T}he $\Xi\Xi$ bond energy, $\Delta B_{\Xi\Xi}$ presented in Table \ref{t02} is connected to the single- and double-$\Xi$ binding energies as
\begin{gather}
 \Delta B_{\Xi\Xi} = B_{\Xi\Xi}(_{\Xi\Xi}^{A}X)-2B_{\Xi}(_{\Xi}^{A-1}X); B_{\Xi\Xi}(_{\Xi\Xi}^{A}X) \nonumber\\ = {[M(^{A-2}X)+ 2M_{\Xi}-M(_{\Xi\Xi}^{A}X)]c^{2}}\label{eq14}
\end{gather}
Filikhin et al., (2017) \cite{filikhin1} reported a single-$\Xi$ separation energy $B_{\Xi}$ = 2.09 MeV and two-$\Xi$ separation energy $B_{\Xi\Xi}$ = 7.635 MeV in $\Xi\Xi\alpha$ three-body system using Differntial Faddeev equations without taking into account Coulomb interaction. Representative plots of the nature of core-$\Xi$ (Eq.~(\ref{eq08})), $\Xi\Xi$ (Eq.~(\ref{eq09})) and $\Xi\Xi$core three-body effective potentials for $_{\Xi\Xi}^6$He and $_{\Xi\Xi}^{16}$C hypernuclei are
shown in Figures~\ref{f02} and \ref{f03} respectively.\\

With gradual increase in the number of contributing partial waves, the two-hyperon separation energy ($B_{\Xi\Xi}$) advances towards convergence which is presented in Table~\ref{t03} and Table~\ref{t04}.

\begin{table}[htbp]
\caption[]{Calculated double-$\Xi$ separation energy $(B_{\Xi\Xi})$, Relative convergence ratio $\frac{\Delta B}{B}$, geometrical size parameters for different $K_{max}$ for $_{\Xi\Xi}^{6}$He system.}\small\smallskip

\tabcolsep=3.6pt
\begin{tabular}{@{}llclllllll@{}}
\hline
\hline
&&&&&&&&&\\[-8pt]
&\textbf{$ B_{\Xi\Xi} $}&&\multicolumn{6}{c}{\textbf{r.m.s radii (fm)}}&\\
\cline{4-9}
$K_{max}$&\textbf{(MeV)}&$\frac{\Delta B}{B}$&$R_{3B}$&$R_{c\Xi}$&$R_{\Xi\Xi}$&$R_{(\Xi\Xi)c}$&$R_{c}^{CM}$&$R_{\Xi}^{CM}$&$\eta$ \\
\hline
&&&&&&&&&\\[-8pt]
0&7.8802&&2.2034&2.7242&3.3103&2.1637&0.8976&2.0839&0.3268\\
4&9.7937&0.2428&2.1330&2.5856&3.1451&2.0524&0.8515&1.9787&0.3531\\
8&10.0112&0.0222&2.1364&2.5922&3.1602&2.0550&0.8525&1.9856&0.3531\\
12&10.1282&0.0117&2.1380&2.5955&3.1653&2.0511&0.8534&1.9884&0.3533\\
16&10.2097&0.0080&2.1370&2.5935&3.1615&2.0561&0.8530&1.9865&0.3542\\
20&10.2709&0.0059&2.1354&2.5904&3.1532&2.0553&0.8527&1.9829&0.3554\\
24&10.3149&0.0043&2.1341&2.5877&3.1432&2.0550&0.8525&1.9796&0.3564\\
\hline
\hline
\end{tabular}
\label{t03}
\end{table}

\begin{figure}[htb]
\begin{minipage}{12.5pc}
\centerline{\includegraphics[width=12.5pc, height=12.5pc]{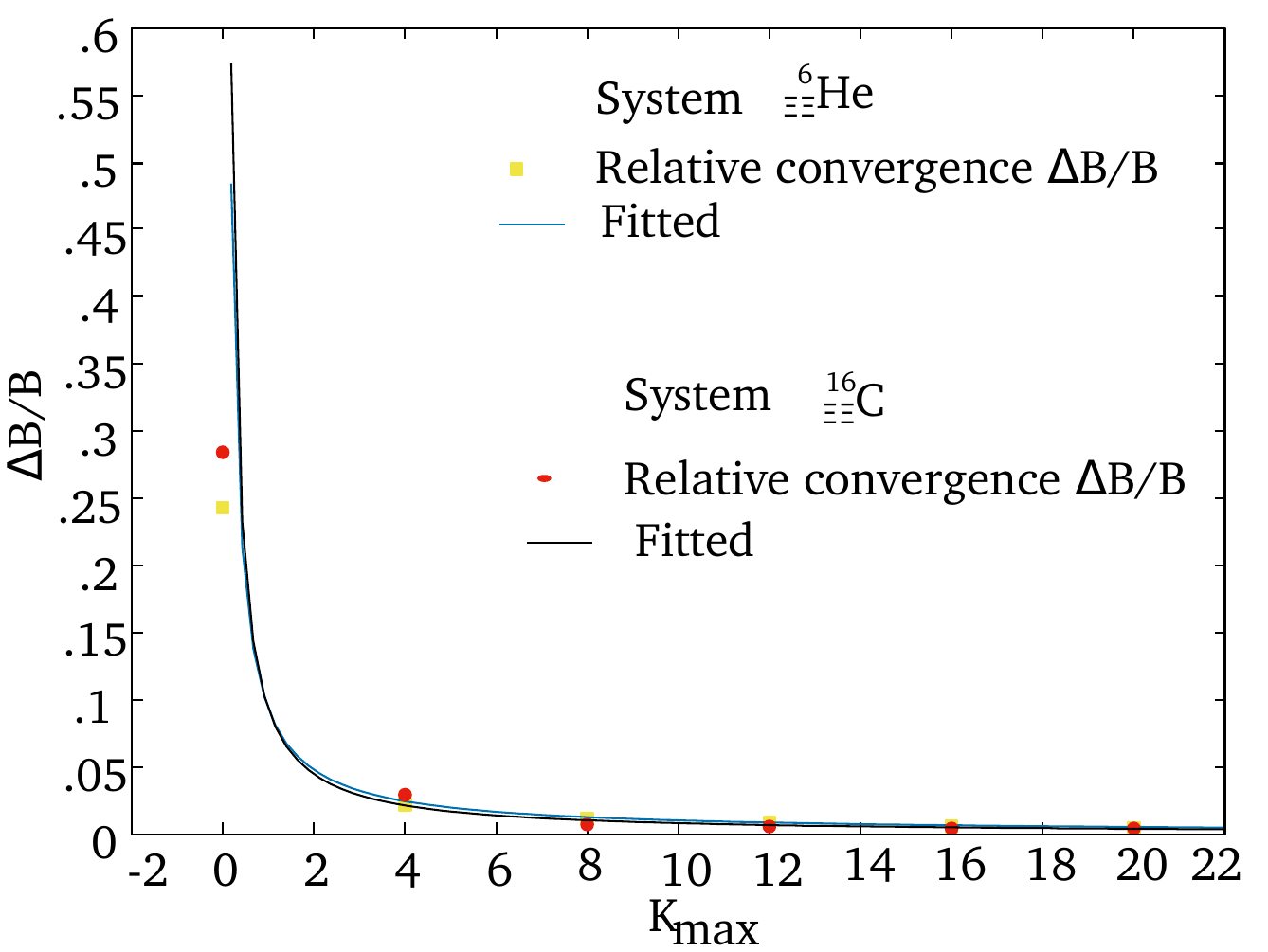}}
\caption{}{Relative convergence ($\frac{\Delta B}{B}$) of double-$\Xi$ separation energy ($B_{\Xi\Xi}$) with respect to $K_{max}$ for $_{\Xi\Xi}^6$He  and $_{\Xi\Xi}^{16}$C using data from Table~\ref{t03} and Table~\ref{t04} respectively.}
\label{f04}
\end{minipage}\hspace{.5pc}
\begin{minipage}{12.5pc}
\centerline{\includegraphics[width=12.5pc, height=12.5pc]{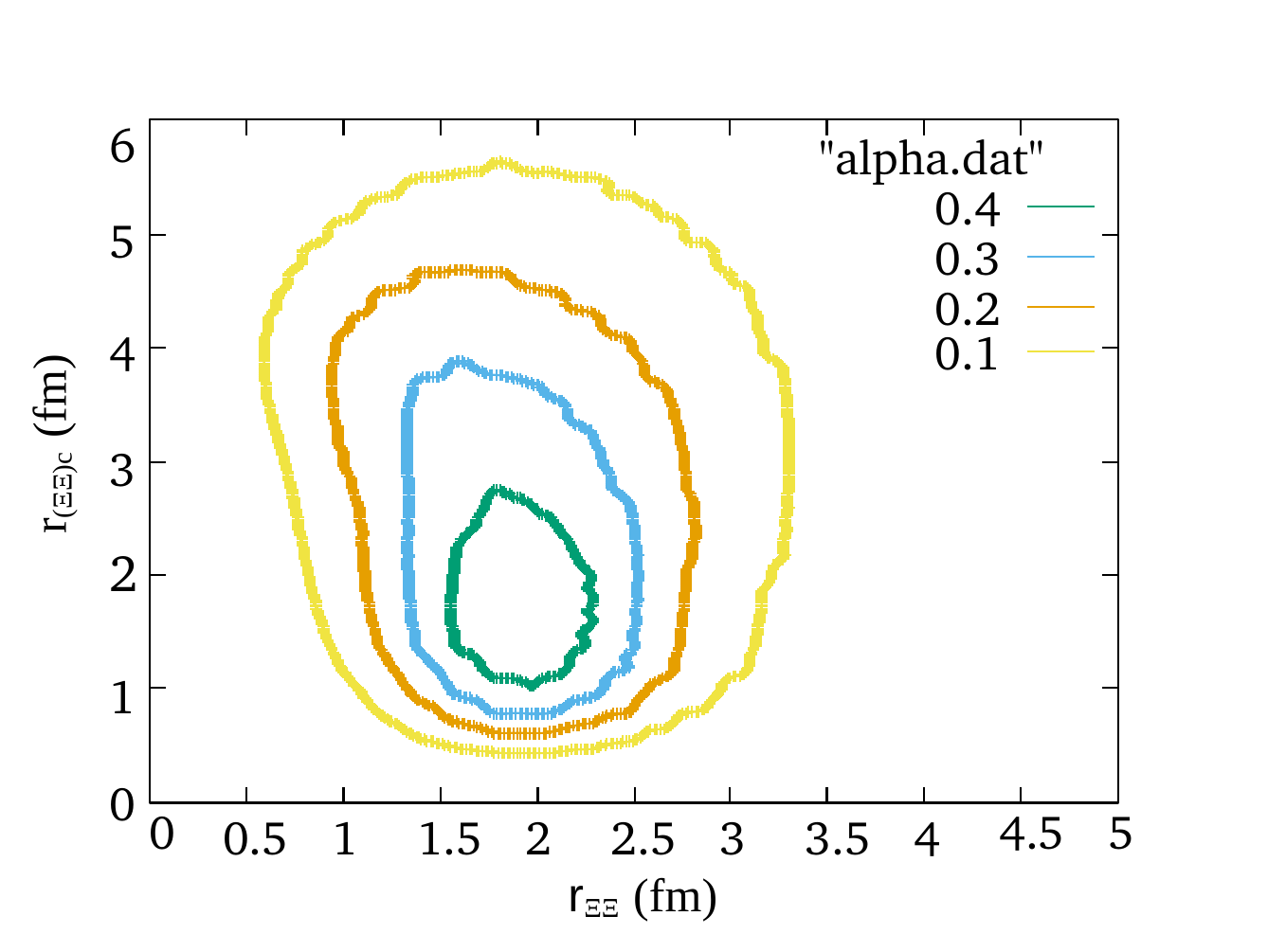}}
\caption{}{Contour plot of probability density distribution of the nucleons and hyperons in the ground state of $_{\Xi\Xi}^6$He system in the $\xi-\chi$ plane (where $\xi$, $\chi$ are in fm).}
\label{f05}
\end{minipage}\hspace{0.5pc}
\end{figure}

\begin{figure}[htb]
\begin{minipage}{12.5pc}
\centerline{\includegraphics[width=12.5pc, height=12.5pc]{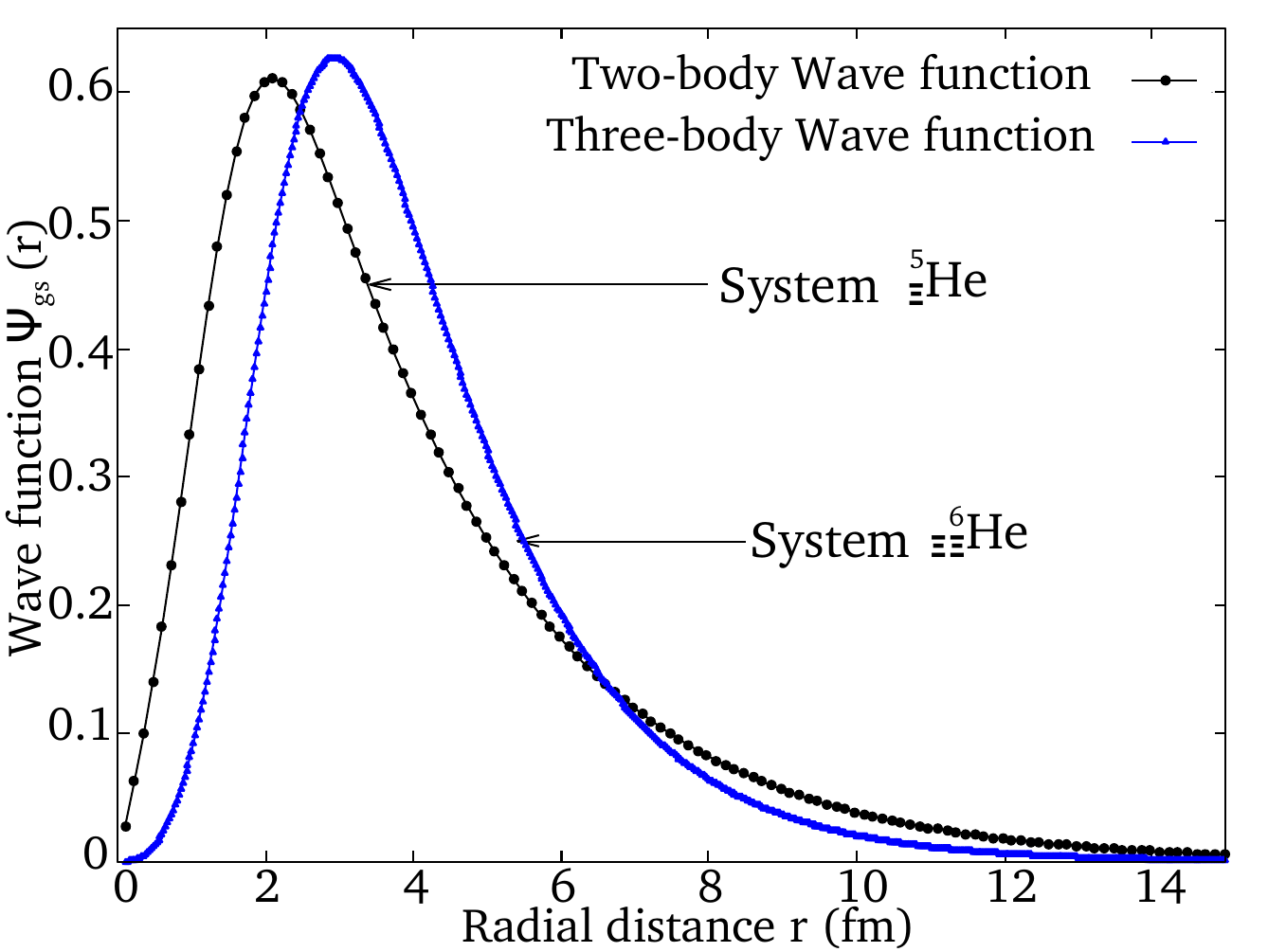}}
\caption{}{Representative plot of the ground state wave function $\psi_{gs}(\rho)$ for two-body $_{\Xi}^5$He and three-body $_{\Xi\Xi}^{6}$He systems.}
\label{f06}
\end{minipage}\hspace{0.5pc}
\begin{minipage}{12.5pc}
\centerline{\includegraphics[width=12.5pc, height=12.5pc]{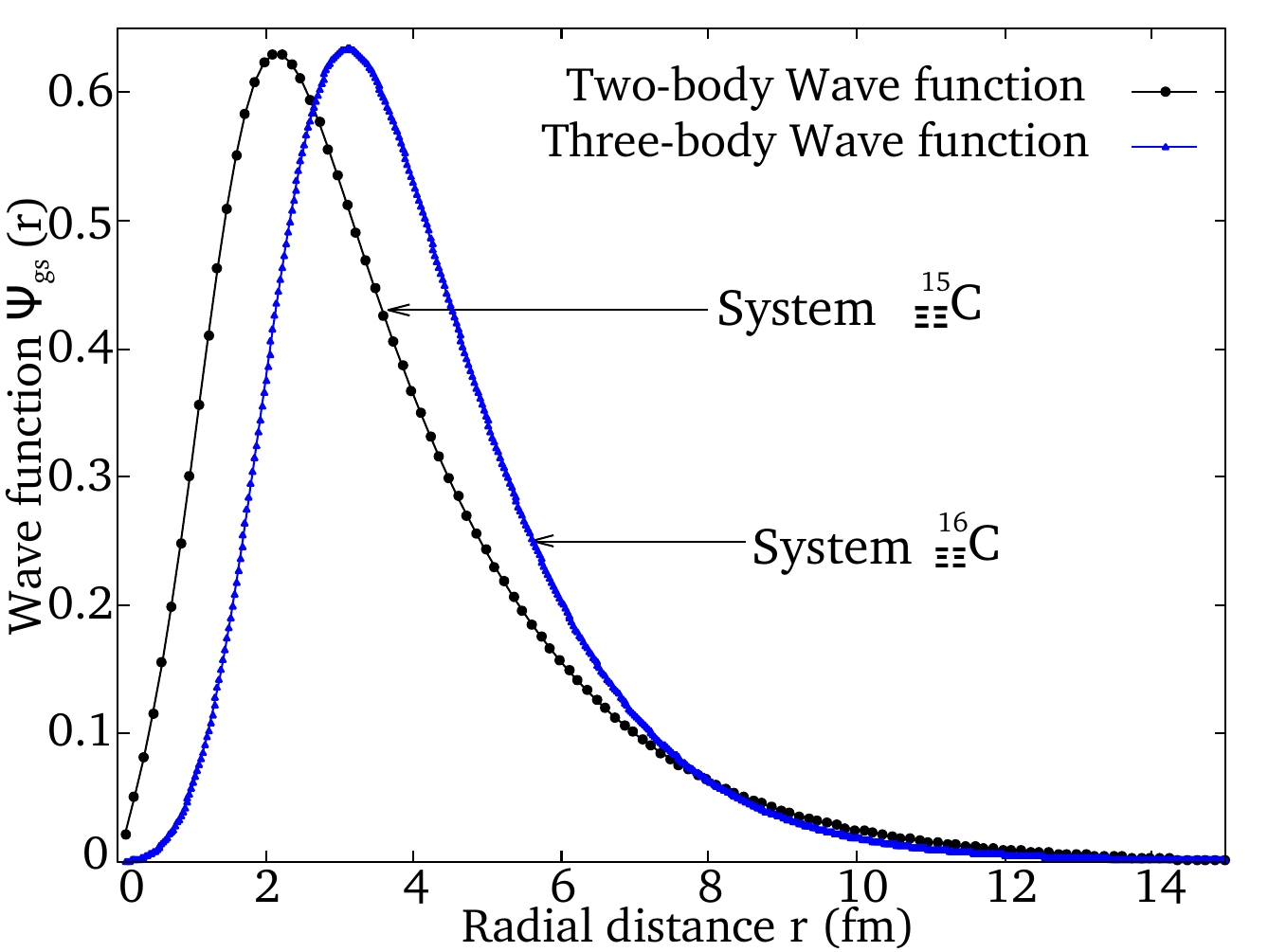}}
\caption{}{Representative plot of the ground state wave function $\psi_{gs}(\rho)$ for two-body $_{\Xi}^{15}$C and $_{\Xi\Xi}^{16}$C systems.}
\label{f07}
\end{minipage}\hspace{0.5pc}
\end{figure}

\begin{table}[htbp]
\caption[]{Calculated double-$\Xi$ separation energy $(B_{\Xi\Xi})$, Relative convergence ratio $\frac{\Delta B}{B}$, geometrical size parameters for different $K_{max}$ for  $_{\Xi\Xi}^{16}$C system.}\small\smallskip

\tabcolsep=3.6pt
\begin{tabular}{@{}llclllllll@{}}
\hline
\hline
&&&&&&&&&\\[-8pt]
&$ B_{\Xi\Xi} $&&\multicolumn{6}{c}{\textbf{r.m.s radii (fm)}}&\\
\cline{4-9}
$K_{max}$&\textbf{(MeV)}&$\frac{\Delta B}{B}$&$R_{3B}$&$R_{c\Xi}$&$R_{\Xi\Xi}$&$R_{(\Xi\Xi)c}$&$R_{c}^{CM}$&$R_{\Xi}^{CM}$&$\eta$ \\
\hline
&&&&&&&&&\\[-8pt]
0&8.6052&&2.7001&2.9310&3.9510&2.1652&0.3627&2.6742&0.2740\\
4&11.0509&0.2848&2.6566&2.6837&3.3711&2.0884&0.3499&2.4215&0.3223\\
8&11.4519&0.0298&2.6581&2.6924&3.3594&2.1041&0.3525&2.4269&0.3269\\
12&11.5339&0.0075&2.6603&2.7052&3.3726&2.1152&0.3544&2.4381&0.3273\\
16&11.5848&0.0048&2.6603&2.7051&3.3714&2.1157&0.3544&2.4379&0.3279\\
20&11.6369&0.0046&2.6596&2.7012&3.3618&2.1145&0.3542&2.4339&0.3289\\
24&11.6832&0.0040&2.6589&2.6970&3.3506&2.1136&0.3541&2.4295&0.3299\\
\hline
\hline
\end{tabular}
\label{t04}
\end{table}
Figure~\ref{f04} is aimed at depicting the pattern of convergence in terms of the ratio $\frac{\Delta B}{B}$ (see Eq.~(\ref{eq15})) with respect to the hyper angular momentum quantum number $K_{max}$ using data presented in Tables~\ref{t03} and \ref{t04}. The relative convergence is defined as
\begin{gather}
\frac{\Delta B}{B}=\frac{B(K+4)-B(K)}{B(K)} \label{eq15}
\end{gather}
and it is found to have an excellent fit with the expression
\begin{gather}
 y(x)=ax^{-b} \label{eq16}
\end{gather}
for a = 0.092974, b = 1.96792 for $_{\Xi\Xi}^6$He and a = 0.092974, b = 1.06679 for $_{\Xi\Xi}^{16}$C respectively.
In Figure~\ref{f05} density distribution of position probability of the hyperons is presented which shows a pear-like shape. A relatively larger separation between the contour lines at the upper region indicates a relatively lower density of hyperons. While a relatively smaller separation between the contour lines at the bottom region of the plot indicates a relatively higher density of hyperons around the nucleus. Thus the plot reflects an interesting fact that the $\Xi$ hyperons are more concentrated in the vicinity of the nucleus.
\noindent{T}he ground state wave function computed at $K_{max} = 24$ for representative cases are shown in Figure~\ref{f06} and Figure~\ref{f07}.
In addition to energy, some geometrical observable of the three-body systems have also been computed using the ground state wave function. These include the root mean squared (r.m.s.) radius of the three-body system defined as
\begin{gather}
R_{3B}=\left[\frac{A_{c}R_{c}^{2}+m_{\Xi}R_{c\Xi}^{2}}{A_{c}+2x}\right]^{1/2} \label{eq17}
\end{gather}
where $A_{c}$, $m_{\Xi}$ are the masses of the core and the hyperon in units of the nucleon mass, $ x=m_{\Xi}/m_{n} $ and $ R_{c}=1.1 A_c^{1/3} fm$ is the matter radius of the core nucleus. The r.m.s. core-$\Xi $ separation is obtained using the expression
\begin{gather}
R_{c\Xi}=\left[<r_{13}^{2}+r_{12}^{2}>/2\right]^{1/2}\label{eq18}
\end{gather}
The calculated r.m.s. radii for increasing  $K_{max}  = 0, 4, 8, ... etc $ up to 24 for $_{\Xi\Xi}^6$He and  $_{\Xi\Xi}^{16}$C are presented in Tables~\ref{t03} and \ref{t04} respectively. A correlation coefficient defined as $\eta = <\frac{r_{(\Xi\Xi)c}^{2}}{\rho^{2}}>$ also computed using the ground state wavefunction. A small value of this coefficient indicates that that the two valence hyperons are positioned on two opposite sides of the core nuclei (i.e., a cigar shape where the $ \Xi $-hyperons are anti-correlated) and a relatively larger value ($\leq$ 1) will suggest the possibility of $ \Xi\Xi $ correlation.

\section{Conclusions} 
In the present scheme, we have first reproduced the observed ground state data for the core-$\Xi$ two-body subsystems of the core-$\Xi\Xi$ three-body systems, by adjusting the parameters of the core-$\Xi$ potential. We then used the same set of core-$\xi$ potential together with the Nijmegen extended softcore $\Xi\Xi$ hyperon-hyperon potential to compute energy and wavefunctions for the corresponding double-$\Xi$ hypernuclei. From the data presented in Tables~\ref{t03} and \ref{t04}, it can be seen that the convergence of energy is slower than that for the geometrical observables for increasing $K_{max}$. The convergence trend of energy with increasing $K_{max}$ has been demonstrated graphically in Figure~\ref{f04}. One can see that for the chosen set of potential $_{\Xi^-}^5$He is bound by about 3.3 MeV over $^5$He which is unbound. And $_{\Xi\Xi}^6$He is bound by 10.3 MeV as compared to $^6$He which is bound by only $0.973 \pm 0.04$
MeV \cite{khan2, zhukov}. The results may also be compared with the corresponding experimental values for $_{\Lambda\Lambda}^6$He, to see that $\Xi$ has more affinity towards nucleons than the $\Lambda$ hyperon since $_{\Lambda}^5$He is bound by about $3.12 \pm 0.02 $ MeV \cite{khan3, takahashi} as compared to 3.3 Mev for $_{\Xi}^5$He and $_{\Lambda\Lambda}^6$He is bound by $7.25 \pm 0.19^{+0.18}_{-0.11}$ MeV \cite{takahashi} or $6.91\pm 0.16$ MeV \cite{ahn2013} as compared to  10.3 MeV for $_{\Xi\Xi}^6$He. Another remarkable point is that absorption of strange hyperon's produce shrinkage effect in target nucleus which is reflected in the calculated RMS matter radius of 2.13 fm for $_{\Xi\Xi}^6$He which is smaller than $2.57\pm 0.1$ fm for $^6$He \cite{khan2, chulkov}. For the single $\Xi$ hypernuclei our calculated data is in fair agreement with the works of others. However, due to the non-availability of observed data on double-$\Xi$ hypernuclei, we cannot compare our results in Table~\ref{t02} with other works. For the available desktop PC configuration, we are allowed to expand the HH basis up to $ K_{max} = 24$ only, which is lagging behind the full convergent solution. But one of the great advantages of the present method is that one can achieve any desired precision in energy by extrapolating the data calculated for smaller $K_{max}$ (Eg. $\leq 24$ in the present case), by applying hyperspherical extrapolation technique discussed in \cite{khan2, schneider}. Finally, we may add that the method adopted here is a robust one for the description of any nuclear or atomic few-body systems subject to the appropriate choice of two-body potentials and symmetry requirements.

\section*{Acknowledgements} 
Authors acknowledge the computational facilities provided by Aliah University, Kolkata, India.\

{\bf Conflict of interest statement:} On behalf of all authors, the corresponding author states that there is no conflict of interest.

\end{document}